Article

# Exploring Avenues beyond Revised DSD Functionals: II. Random-Phase Approximation and Scaled MP3 Corrections

Golokesh Santra, Emmanouil Semidalas, and Jan M. L. Martin*

ACCESS |  Metrics & More |  Article Recommendations |  Supporting Information

**ABSTRACT:** For revDSD double hybrids, the Görling–Levy second-order perturbation theory component is an Achilles' heel when applied to systems with significant near-degeneracy ("static") correlation. We have explored its replacement by the direct random phase approximation (dRPA), inspired by the SCS-dRPA75 functional of Kállay and co-workers. The addition to the final energy of both a D4 empirical dispersion correction and of a semilocal correlation component lead to significant improvements, with DSD-PBEdRPA$_{75}$-D4 approaching the performance of revDSD-PBEP86-D4 and the Berkeley $\omega$B97M(2). This form appears to be fairly insensitive to the choice of the semilocal functional but does exhibit stronger basis set sensitivity than the PT2-based double hybrids (due to much larger prefactors for the nonlocal correlation). As an alternative, we explored adding an MP3-like correction term (in a medium-sized basis set) to a range-separated $\omega$DSD-PBEP86-D4 double hybrid and found it to have significantly lower WTMAD2 (weighted mean absolute deviation) for the large and chemically diverse GMTKN55 benchmark suite; the added computational cost can be mitigated through density fitting techniques.

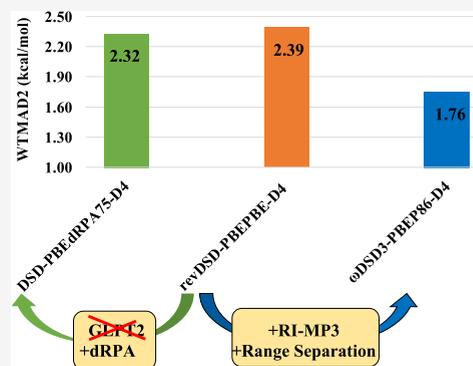

## 1. INTRODUCTION

While the Kohn–Sham density functional theory (KS-DFT)[1] in principle would be exact if the exact exchange-correlation (XC) functional were known, in practice its accuracy is limited by the quality of the approximate XC functional chosen in electronic structure calculations. Over the past few decades, a veritable "zoo" (Perdew's term[2,3]) of such functionals has emerged. Perdew introduced an organizing principle known as the "Jacob's Ladder,"[4] ascending by degrees from the Hartree "vale of tears" (no exchange, no correlation) to the heaven of chemical accuracy: on every degree or rung, a new source of information is introduced. LDA (local density approximation) constitutes the first rung, GGAs (generalized gradient approximations) the second rung, and meta-GGAs (mGGAs, which introduce the density Laplacian or the kinetic energy density) represent the third rung of the ladder. The fourth rung introduces dependence on the occupied Kohn–Sham orbitals: *hybrid* functionals (global, local, and range-separated) are the most important subclass here. Lastly, the fifth rung corresponds to inclusion of virtual orbital information, such as in *double hybrids* (see refs 5–7 for reviews, and most recently ref 8 by the present authors).

Building on the earlier work of Görling and Levy[9] who introduced perturbation theory in a basis of Kohn–Sham orbitals, Grimme's 2006 paper[10] presented the first double hybrid in the current sense of the word. The term refers to the fact that aside from an admixture of (m)GGA and "exact" Hartree–Fock (HF)-like exchange, the correlation is treated as a hybrid of (m)GGA correlation and GLPT2 (second-order

Görling–Levy[9] perturbation theory). Following a Kohn–Sham calculation with a given semilocal XC functional and a given percentage of HF exchange, the total energy is evaluated in the second step as:

$$
\begin{aligned}
E_{DH} = {} & E_{N1e} + c_{X,HF}E_{x,HF} + (1 - c_{X,HF})E_{x,XC} \\
& + c_{C,XC}E_{C,XC} + c_{2ab}E_{2ab} + c_{2ss}E_{2ss} \\
& + E_{disp}[s_6, s_8, s_{ATM}, a_1, a_2, \text{etc}]
\end{aligned} \tag{1}
$$

where $E_{N1e}$ stands for the sum of nuclear repulsion and one-electron energy terms; $E_{X,HF}$ is the HF-exchange energy and $c_{X,HF}$ the corresponding coefficient; $E_{X,XC}$ and $E_{C,XC}$ are the semilocal exchange and correlation energies, respectively; and $c_{C,XC}$ is the fraction of semilocal correlation energy used in the first step. $E_{2ab}$ and $E_{2ss}$ are the opposite-spin and same-spin MP2-like energies obtained in the basis of the KS orbitals from the first step, and $c_{2ab}$ and $c_{2ss}$ are the linear coefficients for the same. Finally, $E_{disp}$ is a dispersion correction, with its own adjustable parameters. As shown, for example, in refs 8, 11 modern double hybrids can achieve accuracies for large, chemically diverse validation benchmarks like GMTKN55[11] (general main-group thermochemistry, kinetics, and noncovalent interactions) that rival those of composite wavefunction theory (cWFT) methods like G4 theory[12,13] (see, however, Semidalas and Martin for some ways to improve cWFT at zero to minimal cost[14,15]).

One Achilles' heel for GLPT2 are molecules with small band gaps (a.k.a absolute near-degeneracy correlation, type A static correlation[16]), owing to the orbital energy difference in the PT2 denominator becoming very small. One potential remedy would be to replace PT2 by the random phase approximation (RPA)[17] for the nonlocal correlation part. From the viewpoint of wavefunction theory, Scuseria and co-workers[18,19] have analytically proven the equivalence of RPA and direct ring coupled clusters with all doubles (drCCD). While the coupled-cluster singles and doubles (CCSD) method is not immune to type A static correlation, it is much more resilient compared to PT2.

The very first foray in this direction was made by Ahnen et al.,[20] who substituted RPA for GLPT2 in the B2PLYP double hybrid.[10] Later, Kállay and co-workers,[21] as well as Grimme and Steinmetz[22] have explored this possibility in greater depth and came up with their own double hybrids featuring the *direct* random phase approximation (dRPA, ref [23] and references therein). The dRPA75 "dual hybrid" of Kállay and co-workers, which uses orbitals evaluated at the PBE75 level (with 75% Hartree−Fock exchange and full PBEc correlation), but only includes pure dRPA correlation in the final energy, is closer in spirit to dRPA than to a double hybrid. In contrast, Grimme and Steinmetz's PWRB95 employs computationally inexpensive mGGA orbitals (specifically, mPW91B95[24,25]) to evaluate a final energy expression consisting of 50% HF exchange, 50% semilocal exchange, 35% dRPA correlation, 71% semilocal correlation, and 65% nonlocal[26] dispersion correction—making it an obvious double hybrid.

One major issue with the dRPA75 was its poor performance for total atomization energies (TAEs, the computational cognates of heats of formation). The authors later remedied that by spin-component scaling:[27] although dRPA is a spin-free method and thus such scaling would have no effect on closed-shell systems, it will affect open-shell cases (most relevantly for TAEs, atoms), particularly as dRPA has a spurious self-correlation energy for unpaired electrons.[28] The so-called SCS-dRPA75 functional employs $c_X = 0.75$, $c_{o−s} = 1.5$, and $c_{s−s} = (2 − c_{o−s}) = 0.5$—addressing the issue for atoms and other open-shell species while being equivalent to dRPA75 for closed-shell species.[27]

In their revision of the S66x8 noncovalent interactions data set,[29] Brauer et al.[30] found that the ostensibly good performance of dRPA75/aug-cc-pVTZ resulted from a spurious error compensation between the basis set superposition error and the absence of a dispersion correction. They also observed, as expected, that the basis set convergence behavior of dRPA is similar to that of CCSD. A D3BJ dispersion correction[31] was parametrized for use with dRPA75 and its parameters found to be very similar to those optimized on top of CCSD (coupled cluster with all singles and doubles[32]); from a symmetry-adapted perturbation theory[33,34] perspective, the most important dispersion term not included in dRPA and CCSD is the fourth-order connected triple excitations term.

In addition, as already mentioned, the dRPA75 and SCS-dRPA75 forms do not include any semilocal correlation contribution in their final energy expressions.

The first research question to be answered in this paper is (see Section 3.1) whether (SCS)dRPA75 can be further improved by not only admitting modern dispersion corrections and semilocal correlation but also reparametrizing against a large and chemically diverse database. The functional form is denoted as DSD-XCdRPAn-Disp, where DSD stands for dispersion-corrected, spin-component-scaled double hybrid, XC stands for the nonlocal exchange-correlation combination used for both the orbital generation in the first step and energy calculation in the second step; $n$ is the percentage of HF-exchange used for both the steps. The final energy for DSD-XCdRPAn-Disp has the form:

$$\begin{aligned}
E_{DH} = {} & E_{N1e} + c_{X,HF}E_{X,HF} + (1 - c_{X,HF})E_{X,XC} \\
& + c_{C,XC}E_{C,XC} + c_{o−s}E_C^{SCS-dRPAo−s} \\
& + c_{s−s}E_C^{SCS-dRPAs−s} + E_{disp}[s_6, s_8, c_{ABC}, a_1, a_2, \text{etc}]
\end{aligned}$$

$$(2)$$

where, $c_{o−s}$ and $c_{s−s}$ stand for opposite-spin and same-spin dRPAc coefficient, respectively. All other terms are the same as eq 1. In this notation, the SCS-dRPA75 dual hybrid is a special case where $c_{X,HF} = 0.75$, $c_{C,XC} = 0$, and $s_6 = s_8 = 0$. As we will show later on, the answer to our research question is affirmative, and the resulting functionals approach the accuracy of the best PT2-based double hybrids known thus far—Mardirossian and Head-Gordon's[35] $\omega$B97M(2) and our own[36] revDSD-PBEP86-D4.

The second research question (to be answered in Section 3.2) is: would taking GLPT2 beyond the second-order improve the performance of revDSD functionals further? Radom and co-workers[37] considered MP3 (third-order many-body perturbation theory), MP4, and CCSD instead of MP2 and found no significant improvement over regular double hybrids. However, this may simply have been an artifact of the modest basis sets and relatively small training set used in ref [37]. Such considerations have been examined in ref [14] where it was also found that the benefits of including an MP3 "middle step" in a 3-tier cWFT can be realized also with a medium-sized basis set for this costly term. In the sections below, we shall consider its addition to global double hybrid revDSD[36] and range-separated $\omega$DSD-type double hybrids using the GMTKN55 data set for training/calibration. Newly developed functionals will be denoted as DSD3 for global DHs and $\omega$DSD3 for range-separated DHs. The final energy expression of a DSD3 functional has the following form:

$$\begin{aligned}
E_{DSD3} = {} & E_{N1e} + c_{X,HF}E_{X,HF} + (1 - c_{X,HF})E_{X,XC} \\
& + c_{C,XC}E_{C,XC} + c_{2ab}E_{2ab} + c_{2ss}E_{2ss} + c_3E_{MP3}^{corr} \\
& + E_{disp}[s_6, s_8, c_{ABC}, a_1, a_2, \text{etc}]
\end{aligned}$$

$$(3)$$

where $E_{MP3}^{corr}$ stands for the MP3 energy component calculated in a basis of HF orbitals, and $c_3$ is a corresponding scaling parameter. All other parameters and energy components are the same as for regular DSD functionals in eq 1. For $\omega$DSD3, the range separation of the HF exchange introduces one additional parameter, the range-separation exponent $\omega$.

We also note that as an alternative to dRPA, GLPT2 might be improved further by energy-dependent regularization methods, as recently introduced by Lee and Head-Gordon.[38] We may explore this possibility in future as a way forward on the PT2-based DSD double hybrids.

**Table 1. Total WTMAD2 (kcal/mol) and Final Parameters for dRPA-Based Dual Hybrids and Their PT2-Based Counterparts[a]**

| functionals | WTMAD2 (kcal/mol) | $c_{X,HF}$ | $c_{X,DFT}$ | $c_{C,DFT}$ | $c_{O-S}$ | $c_{S-S}$ | $s_6$ | $s_8$ | $c_{ATM}$ | $a_1$ | $a_2$ |
|---|---|---|---|---|---|---|---|---|---|---|---|
| SCS-dRPA75 | 4.79 | 0.75 | 0.25 | N/A | 1.5000 | 0.5000 | | | | | |
| optSCS-dRPA75 | 4.71 | 0.75 | 0.25 | N/A | 1.3500 | 0.6500 | | | | | |
| SCS-dRPA75-D3BJ | 2.89 | 0.75 | 0.25 | N/A | 1.5000 | 0.5000 | 0.2528 | [0] | N/A | [0] | 4.5050 |
| optSCS-dRPA75-D3BJ | 2.76 | 0.75 | 0.25 | N/A | 1.3111 | 0.6889 | 0.2546 | [0] | N/A | [0] | 4.5050 |
| DSD-PBEdRPA$_{75}$-D3BJ | 2.38 | 0.75 | 0.25 | 0.1151 | 1.2072 | 0.5250 | 0.3223 | [0] | N/A | [0] | 4.5050 |
| DSD-PBEP86dRPA$_{75}$-D3BJ | 2.36 | 0.75 | 0.25 | 0.1092 | 1.1936 | 0.5268 | 0.3012 | [0] | N/A | [0] | 4.5050 |
| SCS-dRPA75-D4 | 2.83 | 0.75 | 0.25 | N/A | 1.5000 | 0.5000 | 0.3692 | [0] | 0.6180 | −0.0139 | 5.3876 |
| optSCS-dRPA75-D4 | 2.70 | 0.75 | 0.25 | N/A | 1.3100 | 0.6900 | 0.3376 | [0] | 0.4276 | −0.0494 | 5.1979 |
| DSD-PBEdRPA$_{75}$-D4 | 2.35 | 0.75 | 0.25 | 0.1219 | 1.1890 | 0.5281 | 0.3818 | [0] | 0.4571 | −0.2515 | 6.7721 |
| DSD-PBEdRPA$_{75}$-D4 | 2.32 | 0.75 | 0.25 | 0.1339 | 1.1967 | 0.5371 | 0.4257 | [0] | 0.6342 | −0.1455 | 6.3983 |

[a]Constant parameters are in square brackets.

## 2. COMPUTATIONAL METHODS

**2.1. Reference Data.** The primary parametrization and validation set used in this work is the GMTKN55 (general main-group thermochemistry, kinetics, and noncovalent interactions) benchmark[11] by Grimme, and co-workers. This database is an updated and expanded version of its predecessors GMTKN24[39] and GMTKN30.[40] GMTKN55 comprises 55 types of chemical model problems, which can be further classified into five major (top-level) subcategories: thermochemistry of small and medium-sized molecules, barrier heights, large-molecule reactions, intermolecular interactions, and conformer energies (or intramolecular interactions). One full evaluation of the GMTKN55 requires a total of 2459 single point energy calculations, leading to 1499 unique energy differences (complete details of all 55 subsets and original references can be found in Table S1 in the Supporting Information).

The WTMAD2 (weighted mean absolute deviation, type 2) as defined in the GMTKN55 paper[11] has been used as the primary metric of choice throughout the current work:

$$\text{WTMAD2} = \frac{1}{\sum_{i=1}^{55} N_i} \times \sum_{i=1}^{55} N_i \times \frac{56.84 \text{ kcal/mol}}{|\Delta \overline{E}|_i}$$
$$\times \text{MAD}_i \qquad (4)$$

where $|\Delta \overline{E}|_i$ is the mean absolute value of all the reference energies from $i = 1$ to 55, $N_i$ is the number of systems in each subset, MAD$_i$ is the mean absolute difference between the calculated and reference energies for each of the 55 subsets. Mean absolute deviation (MAD) is a more "robust" metric than root-mean-square difference (RMSD), in the statistical sense of the word[41] that it is more resilient to a small number of large outliers than the RMSD. For a normal distribution without systematic errors, RMSD ≈ 5MAD/4.[42]

As one reviewer pointed out, the average absolute reaction energies (AARE) for subsets NBPRC and MB16-43 given in the GMTKN55 paper[11] differ from the corresponding values calculated from the individual data provided in the Supporting Information. If these corrected AARE values were employed in the construction of the WTMAD2 equation, eq 4, then their average, which appears in eq 4 as the overall scale factor, would be 57.76 rather than 56.84. However, as all previously published papers on GMTKN55 (such as refs 3, 8, 31, 36, 43−45) have used the original (smaller) coefficient, we are retaining it as well for the sake of compatibility. This obviously will not affect the ranking between functionals; those who prefer WTMAD2$_{57.76}$ can simply multiply all WTMAD2 values by 1.0162.

Reference geometries were downloaded from the Supporting Information of refs 11 and 46 and used without further geometry optimization.

**2.2. Electronic Structure Calculations.** The MRCC2020[47] program package was used for all calculations involving dRPA correlation. The Weigend–Ahlrichs[48] def2-QZVPP basis set was used for all of the subsets except WATER27, RG18, IL16, G21EA, BH76, BH76RC, and AHB21—where the diffuse-function augmented def2-QZVPPD[49] was employed—and the C60ISO and UPU23 subsets, where we settled for the def2-TZVPP basis set to reduce computational cost.[48] The LD0110-LD0590 angular integration grid was used for all the DFT calculations; this is a pruned Lebedev-type integration grid similar to Grid = UltraFine in Gaussian[50] or SG-3 in Q-Chem.[51]

In their original GMTKN55 paper, Goerigk et al.[11] correlated all electrons in the post-KS steps. However, in a previous study by our group,[36] we have shown that core-valence correlation is best omitted when using the def2-QZVPP basis set (which has no core-valence functions), while a more recent study on composite wavefunction methods indicated that even with correlation consistent core-valence sets, the effect of subvalence electrons on WTMAD2 of GMTKN55 is quite small—benefits gained there are mostly from the added *valence* flexibility of the basis sets.[15] Exceptions were made for MB16-43, HEAVY28, HEAVYSB11, ALK8, CHB6, and ALKBDE10 subsets—where the orbital energy gaps between the halogen and chalcogen valence and metal subvalence shells can drop below 1 hartree, such that subvalence electrons of metal and metalloid atoms must be unfrozen—as well as for the HAL59 and HEAVY28 subsets, where $(n − 1)$spd orbitals on heavy p-block elements were kept unfrozen. We note in passing that, unlike the valence correlation consistent basis sets, the Weigend–Ahlrichs QZVPP basis set is multiple-zeta in the core as well and contains some core-valence polarization functions: see Table 1 of ref 48. At any rate, we have considered[15] the impact of core-valence correlation on GMTKN55 using correlation consistent core-valence basis sets and found (in the context of pure wavefunction calculations) that its impact is on the order of 0.05 kcal/mol—which will be further reduced here through attenuation of the correlation terms.

For the DSD3 and ωDSD3 functionals, QCHEM[51] 5.3 was used throughout. The same "Frozen core" settings and integration grids were applied as were used in the preceding paper on the revDSD and ωDSD functionals.[36] In order to reduce the computational cost, all the MP3 calculations were done using the def2-TZVPP basis set;[48] all other energy components were evaluated using the same basis set combination mentioned above. For technical reasons, HF reference orbitals had to be used for the MP3 steps.

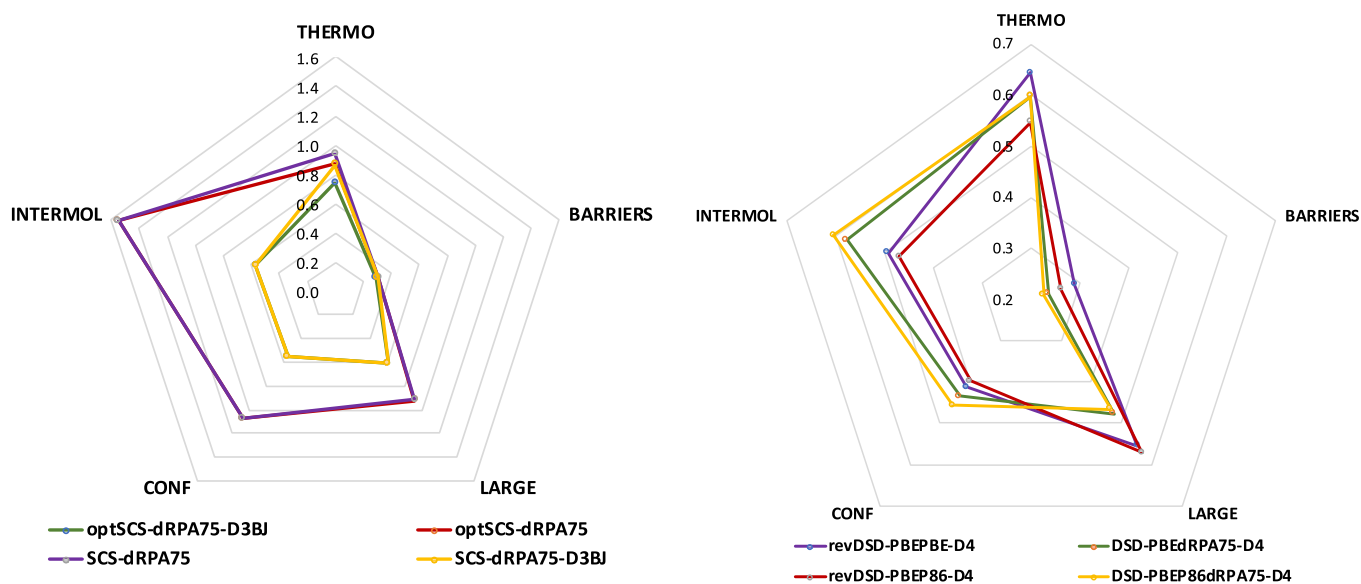

**Figure 1.** Breakdown of total WTMAD2 into five top-level subsets for the dRPA-based dual hybrids (left) and PT2-based vs dRPA-based DSD double hybrids (right) (THERMO, small molecule thermochemistry; BARRIER, barrier heights; LARGE, reaction energies for large systems; CONF, conformer/intramolecular interactions; and INTER, intermolecular interactions). For individual subsets of GMTKN55, see Tables S4–S12 in the Supporting Information.

All the calculations were performed on the ChemFarm HPC cluster in the Faculty of Chemistry at the Weizmann Institute of Science.

**2.3. Optimization of Parameters.** A fully optimized dRPA-based double hybrid will have six empirical parameters: the fraction of global ("exact", HF-like) exchange, $c_{X,HF}$ ($c_{X,DFT} = 1 - c_{X,HF}$); the fraction of semilocal DFT correlation, $c_{C,DFT}$; that of opposite-spin dRPA correlation, $c_{o-s}$; of same-spin dRPA correlation $c_{s-s}$; a prefactor $s_6$ for the D3(BJ) dispersion correction;[31,52,53] and parameter $a_2$ for the D3(BJ) damping function (like in refs 54 and 55 we constrain $a_1 = 0$ and $s_8 = 0$).

However, DSD3-type functionals (see below) introduce one additional parameter ($c_3$) for the MP3 correlation term. For the $\omega$DSD3 family, yet another parameter $\omega$ needs to be considered for range-separation, increasing the total number of empirical parameters to eight—still only half the number involved in the current "best in class" double hybrid $\omega$B97M(2),[35] which has 16 empirical parameters.

We employed Powell's BOBYQA[56] (Bound Optimization BY Quadratic Approximation) derivative-free constrained optimizer, together with scripts and Fortran programs developed inhouse, for the optimization of all parameters.

Once a full set of GMTKN55 calculations is done for one set of fixed nonlinear parameters $c_{X,HF}$ and $c_{C,DFT}$ (for $\omega$DSD3 also $\omega$), the associated optimal values of the *remaining* parameters $\{c_{2ab}, c_{2ss}, (c_3), s_6, a_2\}$ can be obtained in a "microiteration" process. This entire process corresponds to one step in the "macroiterations" in which we minimize WTMAD2 with respect to $\{c_{X,HF}, c_{C,DFT}\}$ and, where applicable, ($\omega$). The process is somewhat akin to microiterations in CASSCF algorithms w.r.t. CI coefficients vs orbitals (see ref 57 and references therein), or QM-MM geometry optimizations where geometric parameters in the MM layer are subjected to microiteration for each change of coordinates in the QM layer (e.g., ref 58).

In view of the small number of adjustable parameters, we have elected, as in our previous studies, to effectively use all of GMTKN55 as both the training and validation set.

## 3. RESULTS AND DISCUSSION

**3.1. GMTKN55 Suite.** In our previous study,[36] we found that refitting the original DSD functionals[54,55] to the large and chemically diverse GMTKN55 data set led to greatly improved performance, particularly for noncovalent interaction and large-molecule reaction energy. Motivated by this prior finding, we attempted first to reoptimize the spin-component-scaling factors in SCS-dRPA75 and obtained WTMAD2 = 4.71 kcal/mol—just a marginal improvement over the original[27] dual hybrid (WTMAD2 = 4.79 kcal/mol).

In the S66x8 noncovalent interaction benchmark paper,[30] dRPA75-D3BJ with basis set extrapolation was found to be the best performer of all DFT functionals. Inspired by this observation, we added a D3BJ correction on top of the Kállay SCS-dRPA75 dual hybrid[27] and found that WTMAD2 dropped from 4.79 to 2.89 kcal/mol. For perspective, it should be pointed out that the lowest WTMAD2 thus far found for a rung-four functional is 3.2 kcal/mol for $\omega$B97M-V.[59] By additionally relaxing the opposite spin and same spin (SS−OS) balance of the dRPA correlation in the optimization, WTMAD2 can be further reduced to 2.76 kcal/mol (see Table 1). As expected, the majority of the improvement comes from the noncovalent interaction and large molecule reaction subsets (Figure 1).

Considering that the energy expression for optSCS-dRPA75-D3BJ contains full dRPA correlation—unlike revDSD double hybrids, where the GLPT2 correlation is scaled down by ∼50%—one can reasonably expect basis set sensitivity. Would improving the basis set beyond def2-QZVPP reduce WTMAD2 further? Extrapolating from def2-TZVPP and def2-QZVPP using the familiar $L^{-3}$ formula of Halkier et al.,[60] we found a reduction by only 0.03 kcal/mol—while using a compromise extrapolation exponent between the $L^{-3}$ for opposite-spin and $L^{-5}$ for same-spin correlation, $\alpha = 3.727$ from solving $[((4/3)^3 -$ $1)^{-1} + ((4/3)^5 - 1)^{-1}]/2 = ((4/3)^\alpha - 1)^{-1}$ reduced WTMAD2 further to 2.70 kcal/mol.

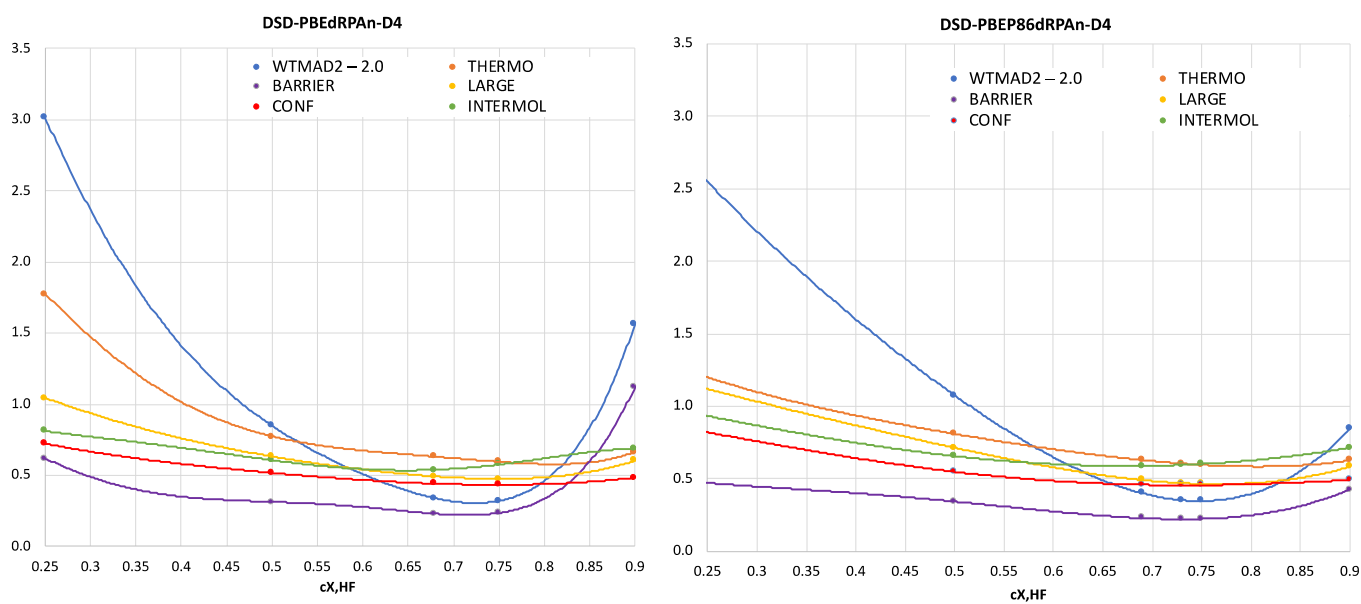

**Figure 2.** Trend of WTMAD2 and top five subcategories with respect to the fraction of HF exchange ($c_{X,HF}$) in DSD-PBEdRPAn-D4 (left) and DSD-PBEP86dRPAn-D4 (right).

What if we "upgrade" D3BJ to the recently published D4[61,62] dispersion term? Aside from the usual four adjustable two-body D4 parameters $s_6$, $s_8$, $a_1$, and $a_2$, the prefactor $c_{ATM}$ of the three-body Axilrod−Teller−Muto term cannot simply be fixed at $c_{ATM}$ = 1 since unlike GLPT2, dRPA does contain $n$-body dispersion.[63,64] Note that when optimized together with the other variables, $s_8$ systematically settled on values near zero; hence, we have constrained $s_8 = 0$ throughout, leaving essentially four dispersion parameters. D4 has thus slightly improved WTMAD2 for SCS-dRPA75 from 2.89 (using D3BJ) to 2.83 kcal/mol. For optSCS-dRPA75, however, it dropped from 2.76 to 2.70 kcal/mol (see Table 1). Among all 55 subsets, BSR36, MCONF, and to some extent WATER27 and PNICO23 benefitted by considering D4. Incidentally, in response to a reviewer query, we have evaluated the impact of the newer revision[65] of D4 (corresponding to version 3 of the standalone dftd4 program) and found the difference for PBE0-D4 to be negligible (0.005 kcal/mol) even for PBE0-D4, where $s_6$ = 1 unlike for the double hybrids at hand.

Thus far, we have only considered dRPA correlation for the nonlocal correlation part of the dual hybrids. Can further improvement be achieved by also mixing some semilocal correlation component into the final energy (i.e., by transforming Kállay's dual hybrid into the true DHDF form)? By doing so, we obtained the DSD-PBEdRPA$_{75}$-D3BJ functional for which WTMAD2 is reduced by an additional 0.38 kcal/mol (see Table 1) at the expense of introducing one additional parameter ($c_{C,DFT}$). The intermolecular interactions subset is the only one that does *not* show a net improvement. The individual data sets that do benefit most are SIE4x4, AMINO20X4, ISOL24, PCONF21, BH76, and PNICO23 (for S66 and BSR36, performance deteriorates). Indeed, this DSD-PBEdRPA$_{75}$-D3BJ (WTMAD2 = 2.36 kcal/mol) compares favorably to its GLPT2-based counterpart, revDSD-PBE-D3BJ (WTMAD2 = 2.67 kcal/mol): a detailed inspection suggests significant improvements for BUT14DIOL, AMINO20x4, TAUT15, HAL59, G21EA, and BHPERI and degradations for SIE4x4 and RG18. If we additionally relax $a_2$ from its fixed value (while keeping $a_1 = s_8 = 0$ fixed) WTMAD2 drops slightly further to 2.33 kcal/mol.

Supplanting D3BJ with the D4[61,62] correction leads to a further drop in WTMAD2 to 2.32 kcal/mol—slightly better than its PT2-based counterpart revDSD-PBEPBE-D4[36] (WTMAD2 = 2.39 kcal/mol). Comparing these two for the five top-level subsets, we found that the dRPA-based double hybrid performs worse for the intermolecular interaction (the lion's share of that due to RG18), comparably for conformer energies, and better for the remaining three (see Figure 1), despite the exception of SIE4x4 due to increased self-interaction error. TAUT15 and G21EA are the two subsets which benefit the most, whereas the two subsets that deteriorate most are SIE4x4 and RG18.

The poor performance of DSD-PBEdRPA$_{75}$-D4 for SIE4x4 can be mitigated by applying the constraints $c_{s-s}$ = 0 and $c_{o-s}$ = 2: MAD for SIE4x4 drops from 9.0 to 4.7 kcal/mol, at the expense of spoiling thermochemical performance.

In a previous study, we found[36] that including the subvalence electron correlation in the GLPT2 step marginally improved WTMAD2 further. This is not the case here: in fact, correlating subvalence electrons with the given basis sets (which do not contain core−valence correlation functions) actually does more harm than good. Therefore, we have not pursued this avenue further (for a detailed discussion and review on basis set convergence for core−valence correlation energies, see ref[66]).

Thus far, we have kept $c_{X,HF}$ fixed at 0.75. What if we include it too in the optimization process? For each value of $c_{X,HF}$, a complete evaluation of the entire GMTKN55 data set is required. We performed such evaluations for five fixed $c_{X,HF}$ points ($c_{X,HF}$ = 0.0, 0.25, 0.50, 0.75, and 0.90), where the same fraction of HF-exchange was used for both the orbital generation and the final energy calculation steps. Interpolation to the aforementioned data points suggests a minimum in WTMAD2 near $c_{X,HF}$ = 0.68; however, upon actual GMTKN55 evaluation at that point, we found that the corresponding WTMAD2 value (2.34 kcal/mol) is very close to the minimum WTMAD2 calculated, 2.32 kcal/mol for $c_{X,HF}$ = 0.75. It thus appears that the WTMAD2 hypersurface in that region is rather flat with respect to variations in $c_{X,HF}$. Performance of the barrier heights subset deteriorates sharply beyond $c_{X,HF} = 0.75$; for all other subsets, however, trends are not as straightforward. Error statistics for conformer energies remain more-or-less unchanged beyond 50% HF exchange. For $c_{X,HF} < 0.5$, a high WTMAD2 value is obtained due to poor performance for small-molecule thermochemistry (see the left side of Figure 2). For each $c_{X,HF}$, the optimized parameters, the WTMAD2, and its breakdown into five top-level subset components can be found in Table S3 in the Supporting Information.

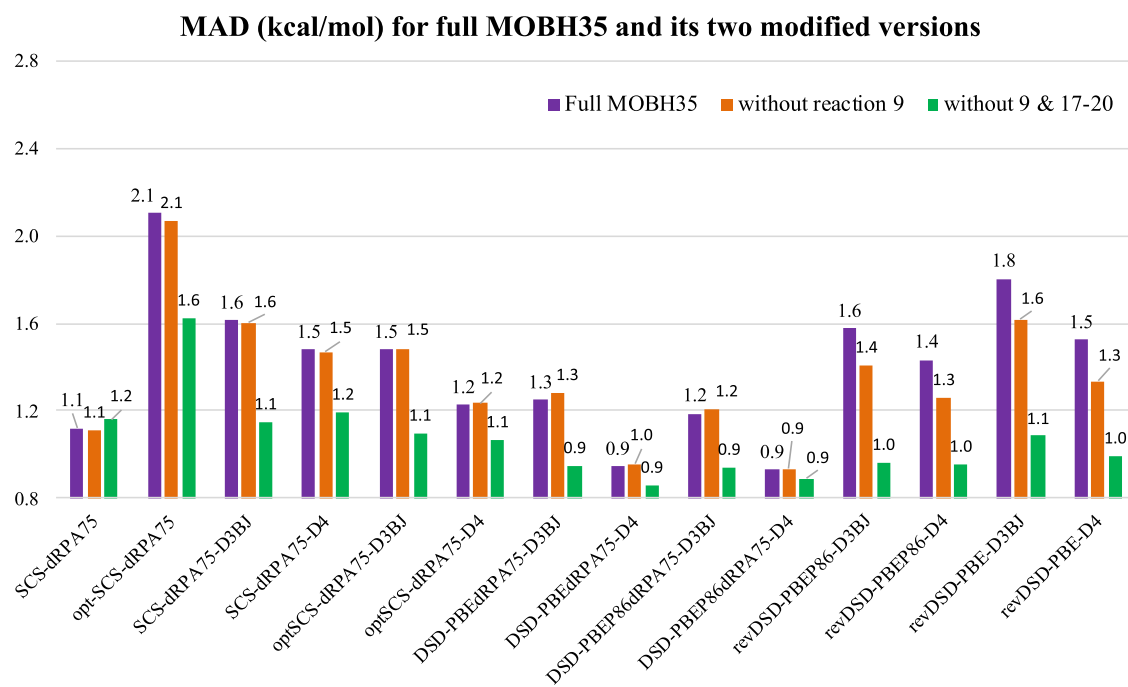

**Figure 3.** MAD (kcal/mol) statistics for the complete and two modified versions of MOBH35.

We also noticed that, with increasing %HF for our functionals, the fraction of DFT correlation in the final energy expression decreases almost linearly and approaches zero near $c_{X,HF} = 0.85$.

For the GLPT2-based double hybrids, we found that in both the original[54,55] and revised[36] parametrizations, the P86c[67,68] semilocal correlation functional yielded superior performance to PBEc[69] (and indeed all other options considered), while we earlier found[54,55] that pretty much any good semilocal exchange functional will perform equally well. Presently, however, we found that DSD-PBEP86dRPA$_n$ alternatives yield only negligible improvements over their DSD-PBEP86dRPA$_n$ counterparts—presumably because the coefficient for the semilocal correlation is so much smaller here.

That being said, our own DSD-PBEdRPA75-D4 and DSD-PBEP86dRPA75-D4 are still inferior to Mardirossian and Head-Gordon's[35] combinatorially optimized range-separated double hybrid, $\omega$B97M(2) (WTMAD2 = 2.13 kcal/mol) (see Table S2 in the Supporting Information). It should be noted here that $\omega$B97M(2) was not trained against GMTKN55 but against a subset of the ca. 5000-point MCGDB84 (main group chemistry data base[70]), although substantial overlap exists between GMTKN55 and MGCDB84.

**3.2. "External" Benchmarks.** Next, we tested our new dRPA-based double hybrids against two separate data sets very different from GMTKN55: the metal–organic barrier height (MOBH35) database by Iron and Janes[71] (see also erratum[72]) and the polypyrroles (extended porphyrins) data set POLY-PYR21.[73,74] Both data sets are known to exhibit moderately strong static correlation (a.k.a., near-degeneracy correlations) effects.[16]

*3.2.1. MOBH35.* This database[71] comprises 35 reactions ranging from $\sigma$-bond metathesis over oxidative addition to ligand dissociations.[71] We extracted the reported best reference energies" from the erratum[72] to the original ref 71 The def2-QZVPP basis set was used for all of our calculations reported here.

Note that these are all closed-shell systems, hence dRPA75, SCS-dRPA75, and optSCS-dRPA75 are equivalent for this problem. Unless a semilocal correlation is introduced into the final energy expression, adding a D3BJ or D4 dispersion correction appears to do more harm than good. However, if the association reactions 17–20 are removed from the statistics, the difference goes away—strongly pointing toward basis set superposition error as the culprit (omitting dispersion corrections would lead to an error cancellation[30]). Among all the functionals tested, DSD-PBEP86dRPA75-D4 and DSD-PBEdRPA75-D4 are the two best performers, both with MAD = 0.9 kcal/mol. Both with D3BJ and D4 corrections, DSD-PBEdRPA75 and DSD-PBEP86dRPA75 are better performers compared to their GLPT2-based revDSD counterparts (the purple bars in Figure 3).

Semidalas et al. (to be published) have recently investigated MOBH35 using a variety of diagnostics for static correlation, as well as recalculated some of the reference energies using canonical CCSD(T) rather than the DLPNO-CCSD(T) approximation.[75] They found that severe type A static correlation in all three structures for reaction 9 (but especially the product) led to a catastrophic breakdown of DLPNO-CCSD(T), to the extent that it can legitimately be asked if even canonical CCSD(T) is adequate. Therefore, omitting this particular reaction and recalculating MADs using the remaining 34 reactions (the orange bars in Figure 3) causes all MADs for the revDSD double hybrids to drop significantly. In contrast, performance for dRPA-based double hybrids remains more or less unchanged. Here too, DSD-PBEdRPA-D4 and DSD-PBEP86dRPA75-D4 are the two best performers.

If, in addition to reaction 9, we also leave out the bimolecular reactions 17–20 (we note that these reactions were omitted from Dohm et al.'s recent revision[76] of MOBH35 as well) and calculate MADs for the remaining 30 reactions (the green bars in Figure 3), the MAD values are seen to drop across the board. However, unlike the full MOBH35, here all the dual hybrids perform similarly, whether we include any dispersion correction or not. The same is true for all the double hybrids. From Figure 3, it is clear that, for DSD-PBEP86dRPA75-D4 and DSD-PBEdRPA75-D4, the MAD values drop slightly compared to the MADs calculated against the original MOBH35.

### 3.2.2. POLYPYR21.
This data set contains 21 structures with Hückel, Möbius, and figure-eight topologies for representative $[4n]$ $\pi$-electron expanded porphyrins, as well as the various transition states between them.[73,74] Among these 21 unique structures, Möbius structures and transition states resembling them exhibit pronounced multireference character (for more details see ref [73]). We have used def2-TZVP basis set throughout; CCSD(T)/CBS reference energies have been extracted from ref [73].

As these are all closed-shell systems, changing the OS-SS balance has no effect on the RMSD value, hence dRPA75, SCS-dRPA75, and optSCS-dRPA75 offer identical error statistics. Adding either D3BJ or D4 dispersion correction on top of that does more harm than good.

Next, similar to what we found for GMTKN55, mixing in semilocal correlation (i.e., DSD-XCdRPAn-Disp) helps quite a bit. Considering the D3BJ dispersion correction, both the dRPA-based double hybrids outperform their PT2-based revDSD counterparts. On the contrary, with D4 dispersion correction, revDSD-D4 functionals have a slight edge over the dRPA-based double hybrids. As expected, the performance variation mainly comes from the Möbius structures, whereas RMSD statistics for the Hückel and twisted-Hückel topologies stay more or less the same for all DSD-DHs (see the third and fourth columns of Table 2).

### 3.3. DSD3 and $\omega$DSD3 Family Functionals: Introducing Scaled Third-Order Correlation.
As mentioned in the Introduction, Radom and co-workers[37] tried to improve on double hybrids by introducing MP3, MP4, and CCSD correlation. Unfortunately, using fairly modest basis sets and fitting correlation energy coefficients to the small and chemically one-sided G2/97[77] database of atomization energies, they failed to discern any significant improvement beyond regular double hybrids. From our previous experience,[36] we know that the use of small, idiosyncratic training sets for empirical functionals may lead to highly suboptimal performance. Thus, here, we are instead employing GMTKN55, which is more than an order of magnitude larger and covers many other types of energetic properties. All the "microiteration" (i.e., linear) parameters were refitted with all $c_{DFT}$, $c_{2ab}$, $c_{2ss}$, and $c_3$; $s_6$ for D3BJ subject to $s_8 = a_1 = 0$, $a_2 = 5.5$ fixed; $s_6$, $a_1$, and $a_2$ for D4 subject to $s_8 = 0$, $c_{ATM} = 1$). Two functionals, DSD-PBEP86 and $\omega$DSD$_{69}$-PBEP86 ($\omega = 0.16$) are considered as the representatives of global and range-separated DHs for the present study. It was previously found,[14] in a cWFT context, that the MP3 term does not change greatly beyond the def2-TZVPP basis set, hence we restrict ourselves to the latter in an attempt to control computational cost.

Total WTMAD2 and optimized parameters for all the DSD3, $\omega$DSD3 and corresponding revDSD functionals are presented in

### Table 2. Mean Absolute Deviations (kcal/mol) and Root Mean Squared Deviations (kcal/mol) for New dRPA-Based DSD-DHs and Original PT2-Based revDSD Functionals on the POLYPYR21 Data Set

| functionals | MAD (kcal/mol) | RMSD (kcal/mol) | | |
| --- | --- | --- | --- | --- |
| | | total | Möbius structures | Hückel and figure-eight structures |
| SCS-dRPA75 | 2.82 | 4.10 | 6.94 | 0.98 |
| optSCS-dRPA75 | 2.82 | 4.10 | 6.94 | 0.98 |
| SCS-dRPA75-D3BJ | 2.88 | 4.18 | 7.09 | 0.96 |
| optSCS-dRPA75-D3BJ | 2.88 | 4.18 | 7.09 | 0.96 |
| DSD-PBEdRPA75-D3BJ | 2.06 | 2.92 | 4.88 | 0.83 |
| DSD-PBEP86dRPA75-D3BJ | 1.96 | 2.78 | 4.64 | 0.79 |
| revDSD-PBEPBE-D3BJ | 2.14 | 3.07 | 5.16 | 0.86 |
| revDSD-PBEP86-D3BJ | 2.07 | 2.94 | 4.94 | 0.80 |
| SCS-dRPA75-D4 | 2.87 | 4.20 | 7.11 | 0.93 |
| optSCS-dRPA75-D4 | 2.89 | 4.23 | 7.18 | 0.92 |
| DSD-PBEdRPA75-D4 | 2.05 | 2.95 | 4.90 | 0.83 |
| DSD-PBEP86dRPA75-D4 | 1.95 | 2.80 | 4.64 | 0.81 |
| revDSD-PBEP86-D4 | 1.93 | 2.87 | 4.78 | 0.82 |
| revDSD-PBEPBE-D4 | 1.90 | 2.81 | 4.66 | 0.84 |

Table 3 (for individual subsets of GMTKN55, see Tables S13–S16 in the Supporting Information). Analyzing the results, we can conclude the following.

- Considering PT2 and MP3 correlation together and scaling the MP3 term by an extra parameter ($c_3$) does improve performance for both the DSD3 and $\omega$DSD3 functionals at the expense of the extra computational cost entailed by the MP3/def2-TZVPP calculations.
- For DSD3 with D4 dispersion correction, the improvement is 0.17 kcal/mol compared to revDSD-PBEP86-D4. Among all 55 individual subsets, the RSE43 subset benefited the most and performance for BHPERI and TAUT15 also improved to some extent. However, for $\omega$DSD3 the performance gain is more pronounced, 0.29 kcal/mol (see Table 3). Inspection of all 55 individual subsets reveals that the RSE43 and TAUT15 subsets showed significant gain in accuracy and AMINO20x4, RG18, ADIM6, and S66 only marginally improved.
- For neither DSD3 nor $\omega$DSD3 can the dispersion correction term be neglected, even if we consider correlation terms beyond PT2.

Using the same GMTKN55 test suite, Semidalas and Martin[14] achieved WTMAD2 = 1.93 kcal/mol for their G4(MP3|KS)-D-v5 cWFT method, which employs the following energy expression,[14]

$$E = E_{HF/CBS} + c_1 E_{MP2|KS, OS/def2-QZVPPD}$$
$$+ c_2 E_{MP2|KS, SS/def2-QZVPPD} + c_3 E_{[MP3-MP2]/def2-TZVPP}$$
$$+ s_6[E(D3BJ)]$$

It differs from the present work in that the semilocal starting point is 100% Hartree–Fock without semilocal correlation, rather than a hybrid GGA as here. Clearly the latter offers an advantage.

Although both the G4(MP3|KS)-D-v5 and DSD3 method use spin component-scaled PT2 correlation and scaled MP3 correlation, the key differences between these two are: no

**Table 3. WTMAD2 (kcal/mol) and all the Optimized Parameters for the Global and Range-Separated DHs with PT2c (revDSD and ωDSD) and PT2c + MP3c (DSD3 and ωDSD3 Functionals)[a,b]**

| functionals | WTMAD2 | ω | $c_{x,HF}$ | $c_{DFT}$ | $c_{2ab}$ | $c_{2ss}$ | $c_3$ | $s_6$ | $s_8$ | $c_{ATM}$ | $a_1$ | $a_2$ |
|---|---|---|---|---|---|---|---|---|---|---|---|---|
| DSD3-PBEP86-D4 | 2.03 | N/A | 0.69 | 0.3784 | 0.6136 | 0.2069 | 0.2443 | 0.6301 | [0] | 1 | 0.3201 | 4.76901 |
| DSD3-PBEP86-D3BJ | 2.12 | N/A | 0.69 | 0.3782 | 0.6085 | 0.2174 | 0.2525 | 0.4582 | [0] | N/A | [0] | [5.5] |
| revDSD-PBEP86-D4 | 2.20 | N/A | 0.69 | 0.4210 | 0.5930 | 0.0608 | [0] | 0.5884 | [0] | 1 | 0.3710 | 4.2014 |
| revDSD-PBEP86-D3BJ | 2.33 | N/A | 0.69 | 0.4316 | 0.5746 | 0.0852 | [0] | 0.4295 | [0] | N/A | [0] | [5.5] |
| DSD3-PBEP86 | 3.34 | N/A | 0.69 | 0.3726 | 0.5402 | 0.5311 | 0.2410 | N/A | N/A | N/A | N/A | N/A |
| ωDSD3-PBEP86-D4 | 1.76 | 0.16 | 0.69 | 0.3048 | 0.6717 | 0.3526 | 0.3057 | 0.5299 | [0] | 1 | 0.0659 | 6.0732 |
| ωDSD3-PBEP86-D3BJ | 1.78 | 0.16 | 0.69 | 0.3063 | 0.6693 | 0.3363 | 0.2842 | 0.3871 | [0] | N/A | [0] | [5.5] |
| ωDSD-PBEP86-D4 | 2.05 | 0.16 | 0.69 | 0.3595 | 0.6610 | 0.1228 | [0] | 0.5080 | [0] | 1 | 0.1545 | 5.1749 |
| ωDSD-PBEP86-D3BJ | 2.08 | 0.16 | 0.69 | 0.3673 | 0.6441 | 0.1490 | [0] | 0.3870 | [0] | N/A | [0] | [5.5] |
| ωDSD-PBEP86 | 2.86 | 0.16 | 0.69 | 0.2749 | 0.6417 | 0.6648 | 0.3620 | N/A | N/A | N/A | N/A | N/A |

[a]Parameters which are kept constant in the optimization cycle are in third bracket. [b]50 systems out of 1499 are omitted: UPU23, C60, 10 largest ISOL24, 3 INV24, and 1 IDISP. N/A, not applicable.

DFT correlation component is present in the final G4(MP3| KS)-D-v5 energy expression, while DSD3 has both scaled HF and DFT exchange, unlike 100% $E_{HF}$ for G4(MP3|KS)-D-v5. Unlike presently, Semidalas and Martin reported[14] that the coefficient for the dispersion term is very small and can be neglected without compromising any significant accuracy (G4(MP3|KS)-D-v6). With a D3BJ dispersion correction ωDSD3-PBEP86 surpasses the accuracy of G4(MP3|KS)-D-v5 method by 0.15 kcal/mol—which can be slightly improved further by considering D4. However, it should be pointed out that DSD3-PBEP86-D3BJ has six adjustable parameters (compared to only four for G4(MP3|KS)-D-v5 and three for G4(MP3|KS)-D-v6), while ωDSD3-PBEP86-D4 has as many as nine.

**3.4. Computational Requirements.** The computational cost of CCSD scales as O($N^6$) with molecular size, and the disk storage scales as O($N^4$). This scaling behavior is similar to that of canonical MP3, which however does not need to store the amplitudes within a direct algorithm. Thus, the estimated speed-up of MP3 over CCSD would be 10−20 times (i.e., the typical number of CCSD iterations). Therefore, in terms of computational cost, this advantage makes the results obtained for DSD3 and ωDSD3 functionals interesting enough without the need for using any further acceleration techniques, such as tensor hypercontraction density fitting (THC-DF-MP3)[78] or the interpolative separable density fitting (ISDF)[79] for the MP3 step.

Following a bug fix to the open-source electronic structure program system PSI4,[80] (version 1.4rc1+) we were able to run RI-MP3 (a.k.a. DF-MP3) for all but a couple dozen of the species for which we had canonical MP3. In the size range of melatonin conformers, we found this to be about seven times faster (wall clock) than conventional MP3, and the overall wall clock time for DSD3-PBEP86-D3BJ and ωDSD3-PBEP86-D3BJ was found to be about three times shorter. It should be noted that our machines are equipped with fast solid state disk scratch arrays with a 3 GB/s bandwidth for sequential writes; for conventional scratch disks, the canonical:RI wall time ratio would be much more lopsided. By way of example, for DSD3-PBEP86-D3BJ, WTMAD2 using conventional MP3 and RI-MP3 components differs by just 0.03 kcal/mol; when substituted for canonical MP3 inside DSD3-PBEP86-D3BJ and ωDSD3-PBEP86-D3BJ, the effects on WTMAD2 are just −0.009 and −0.004 kcal/mol, respectively.

The computational time requirements were checked for two molecules from GMTKN55: one melatonin conformer and one

peptide conformer (see Figure S1 for structures). From Figure 4 we can conclude the following:

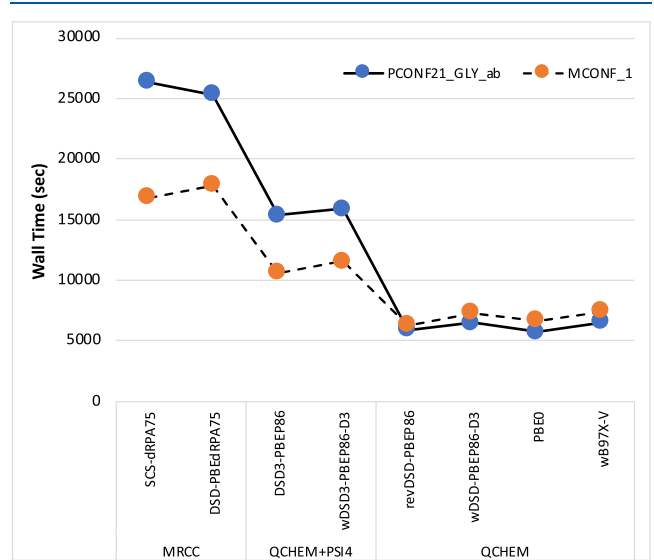

**Figure 4.** Computational time requirements (s) for two systems of GMTKN55 with different hybrid and double hybrid functionals.

(a) Global hybrids and DSD double hybrids (if RI is used), at least in that size range, have broadly comparable computational cost. For very large systems, eventually O($N^5$) will gradually make the RIMP2 the dominant component.

(b) Range-separated hybrids and ωDSD-PT2 again have broadly comparable cost.

(c) With RI-MP3 used, DSD3- and ωDSD3-type functionals cost about 2−3 times as much as an ordinary global- or range-separated double hybrid in this size range.

(d) Our dRPA-based DSD-DHs and Kállay's SCS-dRPA75 cost about 3−5 times as much as global DHs.

## 4. CONCLUSIONS

Analyzing the results presented above for the dRPA-based double hybrids; original and reparametrized form of SCS-dRPA75 dual hybrid; and DSD3- and ωDSD3-type double hybrid functionals (all evaluated against GMTN55), we are able to state the following conclusions. Concerning the first research question:

a) Following the recommendation of Martin and co-workers,[30] adding a dispersion correction on top of the original SCS-dRPA75 significantly improved the WTMAD2 statistics, D4 slightly more so than D3BJ.

b) By additionally admitting a semilocal correlation component into the final energy expression, we were able to obtain DSD-PBEdRPA$_{75}$-D3BJ and DSD-PBEdRPA75-D4 functionals that actually slightly outperform their PT2-based counterparts,[36] revDSD-PBE-D3BJ and revDSD-PBE-D4.

c) We considered different percentages of HF exchange but found the WTMAD2 curve flat enough in the relevant region, for both the DSD-PBEdRPAn-D4 and DSD-PBEP86dRPAn-D4 variants, that $c_{X,HF} = 0.75$ is a reasonable choice.

d) Judging from the SIE4x4 subset, we found that the refitted SS-OS balance in dRPAc apparently causes significant self-interaction errors. This issue can be eliminated by applying the constraint, $c_{s\rightarrow s} = 0$, $c_{o\rightarrow s} = 2$—at the expense of spoiling small-molecule thermochemistry.

Concerning the second research question, we considered a different post-MP2 alternative, namely, the addition of a scaled MP3 correlation term (evaluated in a smaller basis set, and using HF orbitals, for technical reasons). Particularly when using range-separated hybrid GGA orbitals, we achieved a significant improvement in WTMAD2. Especially in conjunction with RI-MP3 or with further acceleration techniques like fragment molecular orbital-based FMO-RI-MP3[81] or the chain-of-spheres approximation for SCS-MP3 as implemented by Izsák and Neese,[82] this approach could potentially be very useful. Head-Gordon and co-workers have very recently shown[83] that the use of DFT orbitals for regular MP3 level calculation results significantly improved performance for thermochemistry, barrier heights, noncovalent interactions, and dipole moments compared to the conventional HF-based MP3. Unlike what Semidalas and Martin[14] observed for their G4(MP3|KS)-D-v5 method, we have found that the dispersion correction term cannot be neglected for DSD3 or ωDSD3 functionals.

More extensive validation calculations of these and prior functionals, both in quantity (using the larger MGCDB84 benchmark[70]) and in system size (MPCONF196,[84] 37CONF8,[85] S30L,[86,87] and to some extent MOR41[88]), are in progress in our laboratory.

## ASSOCIATED CONTENT

### ⓈSupportingInformation

The Supporting Information is available free of charge at https://pubs.acs.org/doi/10.1021/acs.jpca.1c01295.

Abridged details of all 55 subsets of GMTKN55 with proper references; breakdown of total WTMAD2 into five major subcategories for the original and refitted SCS-DRPA75 variants, DSD-PBEdRPA$_{75}$, DSD-PBEP86dR-PA$_{75}$, and corresponding revDSD functionals with dispersion correction; optimized parameters and breakdown of total WTMAD2 into five top-level subsets for DSD-PBEdRPAn and DSD-PBEP86dRPAn both with D3BJ and D4 dispersion correction; MAD, MSD, and RMSD as well as breakdown of total WTMAD2 by each subset for each new functionals we propose here; figure showing the structures of two systems upon which we experimented the computational time requirements for different functionals; and ORCA sample inputs for revDSD-PBEP86-D3BJ and revDSD-PBEPBE-D3BJ (PDF)

## AUTHOR INFORMATION

### Corresponding Author

Jan M. L. Martin − Department of Organic Chemistry, Weizmann Institute of Science, 7610001 Reḥovot, Israel; ⓞ orcid.org/0000-0002-0005-5074; Email: gershom@weizmann.ac.il

### Authors

Golokesh Santra − Department of Organic Chemistry, Weizmann Institute of Science, 7610001 Reḥovot, Israel; ⓞ orcid.org/0000-0002-7297-8767

Emmanouil Semidalas − Department of Organic Chemistry, Weizmann Institute of Science, 7610001 Reḥovot, Israel; ⓞ orcid.org/0000-0002-4464-4057

Complete contact information is available at: https://pubs.acs.org/10.1021/acs.jpca.1c01295

### Notes

The authors declare no competing financial interest.

## ACKNOWLEDGMENTS

We would like to acknowledge helpful discussions with Drs. Mark A. Iron and Irena Efremenko (WIS), and Prof. Mercedes Alonso Giner (Free University of Brussels). Mr. Minsik Cho (Brown University) is thanked for parallelizing the parameter optimization process while he was a Kupcynet-Getz International Summer School intern in our group. We also thank Mr. Nitai Sylvetsky for critical comments on the draft manuscript. G.S. acknowledges a doctoral fellowship from the Feinberg Graduate School (WIS). This research was funded by the Israel Science Foundation (grant 1969/20) and by the Minerva Foundation (grant 20/05). The work of E.S. on this scientific paper was supported by the Onassis Foundation—Scholarship ID: F ZP 052-1/2019-2020.